\documentclass[fleqn,twoside]{article}
\usepackage[headings]{espcrc2}

\readRCS
$Id: espcrc2.tex,v 1.2 2004/02/24 11:22:11 spepping Exp $
\usepackage{graphicx}
\usepackage[figuresright]{rotating}
\newfont{\cmsy}{cmsy12}

\newcommand{\AmS}{{\protect\the\textfont2
  A\kern-.1667em\lower.5ex\hbox{M}\kern-.125emS}}

\title{$\eta$ and $\eta$' Physics at MAMI}

\author{M. Unverzagt\address[Mainz]{Institut f\"ur Kernphysik, 
        Johannes Gutenberg-Universit\"at Mainz, \\ 
        D-55099 Mainz, Germany}%
        }

\runtitle{$\eta$ and $\eta'$ Physics at MAMI}
\runauthor{M. Unverzagt}

\begin{document}

\begin{abstract}
The Crystal Ball at MAMI setup offers an excellent possibility to study decays of the $\eta$ and $\eta'$ meson. Here, recent results of the Crystal Ball at MAMI experiment from $\eta$ meson decays are presented. Furthermore, future perspectives of this experiment in the field of $\eta$ and $\eta'$ physics are described.

\vspace{1pc}
\end{abstract}

\maketitle

\section{Introduction}

The decays of the $\eta$ and $\eta'$ mesons provide unique information on the understanding of low-energy Quantum Chromo Dynamics (QCD), the field theory of strong interaction. Since perturbative QCD cannot be applied in the low-energy region, because the strong coupling constant $\alpha_s$ is large, other methods like lattice QCD, chiral perturbation theory ($\chi$PT) or model-dependent approaches are used. These can be tested by studying the different decay modes of the $\eta$ and $\eta'$ mesons. Especially, the decays of the $\eta'$ can be used to probe the applicability of $\chi$PT (the $\eta'$ mass is of the order of the chiral symmetry breaking scale of $4\pi f_\pi \simeq 1.2$\,GeV). Moreover, many models and theories of hadron interaction can be tested. One can also search for the violation of lepton--family number and place limits on the masses and couplings of many proposed lepto--quark families, see refs~\cite{Goi02,Ram02,Mar05,Nic02,Her95}. Decays of the $\eta$ and $\eta'$ are also suitable to search for violations of $C$, $CP$, and $CPT$ invariance~\cite{Nef95}.

The high potential of this field is reflected by the number of experiments that have been or are currently active in $\eta$ and $\eta'$ physics. At the Mainz Microtron (MAMI) with the maximum beam energy of 1604 MeV and the Crystal Ball/TAPS setup~\cite{Ore82,Sta01,Nov91,Gab94}, a perfect environment for $\eta$ and $\eta'$ studies is given. The high statistics achievable at MAMI together with the good signal-to-background conditions makes investigations of the $\eta$ and $\eta'$ mesons especially competitive at MAMI. Other experiments doing research in this field are the KLOE experiment at the DA$\Phi$NE collider, the Crystal Barrel detector at ELSA and WASA at COSY. Also other experiments, like BES III, CLAS and GAMS, are capable of measuring $\eta$ and $\eta'$ decays, but the capabilities at MAMI--C have the potential to make MAMI the leading research laboratory for $\eta$ and $\eta'$ mesons, at least for neutral decays.

\section{Experimental Setup}

The electron accelerator MAMI consists of a cascade of three Race-Track-Microtrons (RTM)~\cite{Her83}. Recently a fourth stage, a Harmonic-Double-Sided Microtron (HDSM)~\cite{Kai08}, has been finished. MAMI provides a very stable electron beam (energy drift $\delta E < 100$~keV) with a maximum energy of $E_0 = 1558$~MeV and an energy width of $\sigma_E/E < 3 \cdot 10^{-5}$. High currents (110~$\mu$A) with polarisations up to 80~\% can be achieved. Very recently a first test was successful to operate MAMI--C with an electron energy of $E_0 = 1604$~MeV, which is especially important for the production of the $\eta'$ meson as discussed below.

At the MAMI complex three major experiments are installed: the A1 experiment with four high resolution spectrometers for electron scattering measurements, the A2--setup with the Crystal Ball and TAPS detectors, as well as the A4 high-rate calorimetric detector for the measurement of parity violation in elastic ep--scattering. The MAMI accelerator is typically in operation for about 7000 hours per year, which proves the high interest in hadron physics experiments at MAMI.

\begin{figure}
        \begin{centering}
        \includegraphics[scale=0.58]{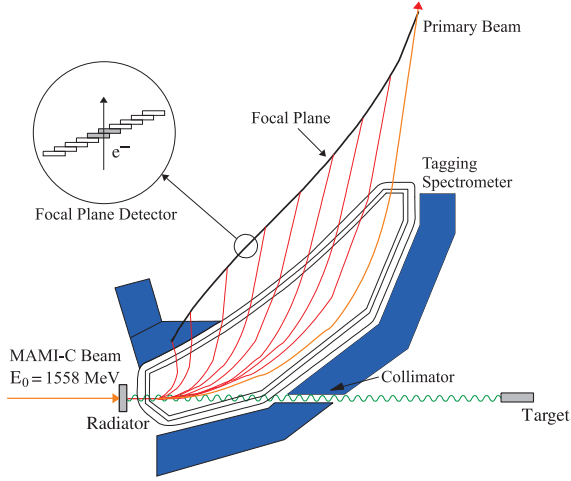}
        \par\end{centering}
        \caption{The Glasgow photon tagging spectrometer at MAMI.}
        \label{pstagger}
\end{figure}

Within the A2 collaboration, experiments are performed with a real photon beam. The photon beam for the production of the $\eta$ and $\eta'$ mesons is derived from the production of Bremsstrahlung photons during the passage of the MAMI electron beam through a thin radiator. The Glasgow photon tagging spectro\-meter~\cite{Ant91,Hal96} at Mainz (fig.~\ref{pstagger}) provides energy tagging of the photons, through $E_\gamma = E_0 - E_{e^-}$, by detecting the post--radiating electron energy. It can determine the photon energy with a resolution of 2 to 4\,MeV depending on the incident beam energy, with a single--counter time resolution $\sigma_{t}=0.17$\,ns~\cite{McG08}. Each counter can operate reliably to a rate of $\sim 1$\,MHz, corresponding to a photon flux of $2.5 \cdot 10^{5}\,\mathrm{(s\,MeV)}^{-1}$. Photons can be tagged in the energy range from 5 to 93\,\% of the incoming electron energy $E_0$. This means that the highest tagged photon energy, at an electron beam energy of $E_0 = 1558$\,MeV, is $E_\gamma \approx 1448$\,MeV.

\begin{figure}
        \begin{centering}
        \includegraphics[scale=0.60]{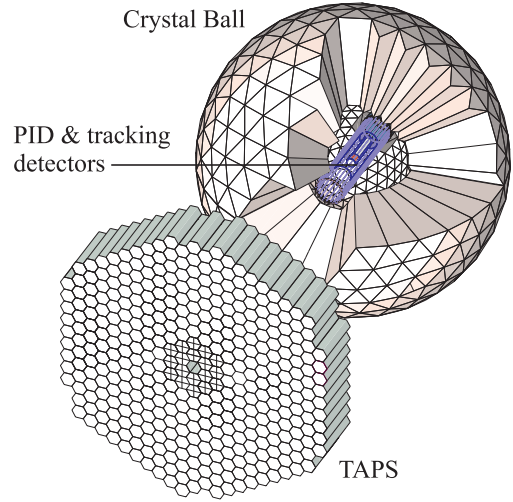}
        \par\end{centering}
        \caption{The detector setup: the Crystal Ball calorimeter with cut--away section showing the inner detectors and the TAPS forward wall. The inner two rings of TAPS are made of PbWO$_4$ crystals.}
        \label{psCBTAPS}
\end{figure}

The central detector system, surrounding the liquid hydrogen target (see fig.~\ref{psCBTAPS}), consists of the Crystal Ball calorimeter (CB)~\cite{Ore82,Sta01} combined with a barrel of scintillation counters for particle identification. The CB is a highly segmented 672-element NaI(Tl), self triggering photon spectrometer constructed at SLAC in the 1970's. Each element is a 41 cm (15.7 radiation lengths) long truncated triangular pyramid. They are all read out by individual photomultipliers. The CB has an energy resolution of $\Delta E/E=0.020(E[\mathrm{GeV}])^{0.36}$, angular resolutions $\sigma_{\theta}$ of $2-3^{\circ}$ and $\sigma_{\phi}$ of $\sigma_{\theta}/\sin\theta$ for electromagnetic showers. The Crystal Ball system is equipped with a barrel detector of 24 scintillator strips (50 mm length, 4 mm thickness at a radius of 15~cm around the photon beam line), which makes particle identification available (PID).

At forward polar angles $<21^{\circ}$ the TAPS detector~\cite{Nov91,Gab94} provides acceptance for particle detection. Hence, the full detector system, as shown in fig.~\ref{psCBTAPS}, is almost hermetic. The TAPS forward wall is composed of 384 BaF$_{2}$ elements, each 25\,cm in length (12 radiation lengths) and hexagonal in cross section, with a diameter of 59\,mm. Every TAPS element is covered by a 5 mm thick plastic veto scintillator. The single counter time resolution is $\sigma_{t}=0.2$\,ns. The energy resolution can be described by $\Delta E/E=0.018+0.008/(E[\mathrm{GeV}])^{0.5}$. The angular resolution in the polar angle is better than $1^{\circ}$, and in the azimuthal angle it improves with increasing $\theta$, being always better than 1/R radian, where R is the distance in centimeters from the central point of the TAPS wall surface to the point on the surface where the particle trajectory meets the detector. The 2 inner rings of 18  BaF$_{2}$ elements have recently been replaced by 72 PbWO$_4$ crystals each 20\,cm in length (22 radiation lengths), see fig.~\ref{psCBTAPS}. The higher granularity improves the rate capability as well as the angular resolution. The energy resolution of PbWO$_4$ for photons is similar to BaF$_2$ at room temperature~\cite{Nov02}.

\section{$\eta$ Physics}\label{hdetaphysics}

Since the installation of the Crystal Ball at MAMI in 2003, several measurements on the $\eta$ have been performed. In the following section \ref{hdeta} the determination of the $\eta$ mass, of the Dalitz plot parameter $\alpha$, and the branching ratio of $\eta \to \pi^0 \gamma \gamma$ are shown. In sect.~\ref{hdetafut} future investigations of the $\eta$ meson are described.

\subsection{Recent results}\label{hdeta}

The mass of the $\eta$ meson has been discussed controversially in recent years. Before 2000, three different experiments~\cite{Dua74,Plo92,Kru95} yielded comparable masses of the $\eta$ meson. The PDG then used these results to calculate a weighted mean mass $m_{\eta} = (547.30 \pm 0.12)$\,MeV~\cite{Gro00}. In 2002, the NA48 collaboration published~\cite{Lai02} a very precise result, $m_{\eta} = (547.84 \pm 0.05)$\,MeV, which deviated significantly from the world average reported by the PDG. This created the motivation to repeat the previous Mainz measurement at MAMI~\cite{Kru95}, especially after another precise measurement by the GEM collaboration at the COSY facility~\cite{Abd05} gave the result $m_{\eta} = (547.31\pm 0.04)$\,MeV, in agreement with the old measurements of the $\eta$ meson mass.

\begin{figure}
        \begin{centering}
        \includegraphics[scale=0.40]{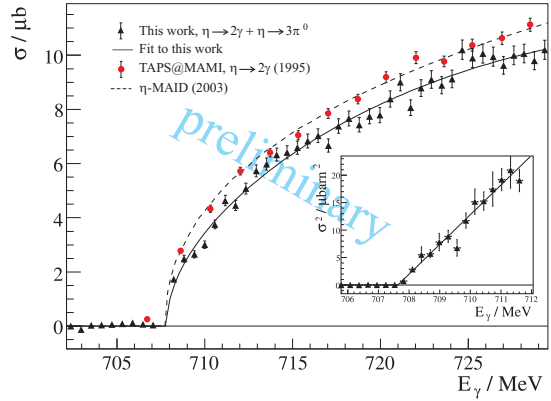}
        \par\end{centering}
        \caption{Total cross section for the $\gamma p \to p \eta$ reaction compared to the measurement made with TAPS at MAMI \cite{Kru95} and the prediction of the $\eta$-MAID partial wave analysis~\cite{Chi03}. The solid line represents a fit to the results of Crystal Ball at MAMI. The inset shows the linear behavior of $\sigma^2$ near threshold because of the strong coupling to the S$_{11}$ resonance.}
        \label{psetamass}
\end{figure}

The $\eta$ mass was determined by measuring the threshold of the reaction $\gamma p \to p \eta$~\cite{Nik09}. The tagger microscope~\cite{Rei06}, a small focal plane detector with a much higher resolution than the main focal plane spectrometer of the tagger, was used for the first time to get the total cross section of the reaction (see fig.~\ref{psetamass}). Then the $\eta$ mass was determined from the photoproduction threshold $E_{thr}$ by
\begin{equation}
 	m_{\eta} = -m_p + \sqrt{m_p^2 + 2 E_{thr} m_p}
\end{equation}

The preliminary MAMI result, $m_\eta = (547.76 \pm 0.01_{stat} \pm 0.07_{syst})$\,MeV, is in agreement with the most recent $\eta$ mass measurements by the NA48~\cite{Lai02}, KLOE~\cite{Amb07} and CLEO~\cite{Mil07} collaborations, though it disagrees with the GEM collaboration result~\cite{Abd05}. The uncertainty for the new $\eta$ mass measurement has been improved in comparison to the previous Mainz experiment by a factor of $\sim$\,$3$, but the two results disagree with each other by two $\sigma$. As reason for this disagreement a much better tagger energy calibration in the new measurement and the higher resolution of the tagger microscope were found.

At MAMI two new determinations of the Dalitz plot parameter, $\alpha$, of the decay $\eta \to 3\pi^0$ were performed. The $\eta \rightarrow 3\pi^0$ decay violates isospin symmetry. Therefore, it offers a unique possibility to study symmetries and symmetry-breaking characteristics of strong interactions. Because electromagnetic contributions to the amplitude can be neglected \cite{Sut66,Bau96,Dit08}, this decay occurs due to the isospin breaking part of the QCD Hamiltonian:
\begin{equation}
        \mbox{\cmsy H \usefont{T1}{ptm}{m}{n}}_{\not{\:\mathrm{I}}} = \frac{1}{2}(m_{\mathrm{u}}-m_{\mathrm{d}})(\bar{\mathrm{u}}\mathrm{u} - \bar{\mathrm{d}}\mathrm{d}).
        \label{eqIsoBreak}
\end{equation}
Hence, the amplitude is proportional to the mass difference $m_\mathrm{u} - m_\mathrm{d}$ of the two lightest quarks u and d. Calculations of the decay amplitude are usually based on the framework of $\chi$PT. Thus, this measurement provides also a very important test for $\chi$PT.

\begin{figure}
        \begin{center}
                \includegraphics[scale=0.37]{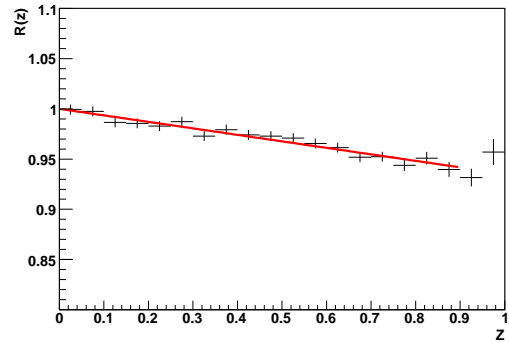}
                \caption{$\alpha$ was obtained directly from the fit of the function $c(1+2 \alpha z)$. In the figure, $R(z)$ has been scaled with $c$ so that the fitted line has an intercept equal to 1 on the $R(z)$-axis.}
                \label{psSlope}
        \end{center}
\end{figure}

The squared absolute value of the decay amplitude may be expanded around the centre of the Dalitz plot:
\begin{equation}
        R(z) = |A(\eta \rightarrow 3\pi^0)|^2 = |N|^2 (1 + 2 \alpha z + \ldots),
        \label{eqAmpExpand}
\end{equation}
where $N$ is a normalisation constant, which is equal to the amplitude that would apply, if the decay proceeded only according to the available phase space. The Dalitz plot parameter, $\alpha$, describes the pion energy dependence of the squared absolute value of the decay amplitude up to first order of the expansion. The parameter $z$ is given by
\begin{equation}
        z = 6 \sum_{i=1}^3 \left( \frac{E_i-m_{\eta}/3}{m_{\eta}-3m_{\pi^0}}\right)^2.
        \label{eqz}
\end{equation}
Here $E_i$ represents the pion energies in the $\eta$ rest frame. Figure~\ref{psSlope} shows the extraction of $\alpha$ from $R(z)$.

The two MAMI results for $\alpha$ were obtained from different data sets with different electron beam energies, $E_{e^-} = 883$\,MeV~\cite{Unv09} and $E_{e^-} = 1508$\,MeV~\cite{Pra09}. Thus, they can be considered as independent measurements. These results, with $1.8 \cdot 10^6$ and $3 \cdot 10^6$ events, represent the two world highest statistics for the $\eta \to 3\pi^0$ decay. The two totally independent values for $\alpha$ from MAMI agree perfectly with each other and are consistent with other experiments. Pure $\chi$PT calculations give the wrong sign for $\alpha$, which is fixed by the experiments to be negative. Dispersion calculations give the correct sign, but differ in the absolute value by several standard deviations. Only the UCHPT approach~\cite{Bor05} produces a Dalitz plot parameter consistent with the experiments.

A cusp effect due to the $\pi^+ \pi^- \to \pi^0 \pi^0$ rescattering reaction, similar to the one observed by the NA48 collaboration in $K^+ \to \pi^+ \pi^0 \pi^0$~\cite{Giu09,Bat06}, should also appear in the $\eta \to 3\pi^0$ decay. This charge exchange reaction also has an impact on the Dalitz plot parameter. According to~\cite{Dit08} a 5\,\% effect on $\alpha$ should be visible. The two recent MAMI measurements~\cite{Unv09,Pra09} with the given statistics did not find a significant indication for a cusp effect.

The rare, double--radiative decay $\eta \to \pi^0 \gamma \gamma$ has attracted much attention as there are large uncertainties on its experimental decay width and in its theoretical predictions. The uncertainties in $\chi$PT calculations of the amplitude for the $\eta \to \pi^0 \gamma \gamma$ transition are related to the fact that the leading term $O(p^2)$ and the tree contributions at $O(p^4)$ are zero as neither $\pi^0$ nor $\eta$ can emit a photon. The pion and kaon loops at $O(p^4)$ are greatly suppressed due to G--parity invariance and the large mass of the kaons, respectively. The main contribution to the $\eta \to \pi^0 \gamma \gamma$ decay amplitude comes from the $O(p^6)$ counterterms that are needed in $\chi$PT to cancel various divergences. The coefficients of these counterterms are not determined by $\chi$PT itself; they depend on the model used for the calculation. As the $\eta \to \pi^0 \gamma \gamma$ decay has a three--body final state, its Dalitz plot reflects the decay amplitude. Thus, a complete test of $\chi$PT with its $O(p^6)$ chiral coefficients requires an experimental measurement of both the $\eta \to \pi^0 \gamma \gamma$ decay rate and the Dalitz plot.

One can see that the existing experimental results and theoretical calculations for $\eta \to \pi^0 \gamma \gamma$ vary a lot. Also, strictly speaking, the agreement between the measured and calculated decay width is not sufficient to prove $\chi$PT calculations. Since every calculation of $\eta \to \pi^0 \gamma \gamma$ makes a specific prediction for the decay Dalitz plot, the experimental measurement of this plot must confirm the theoretical prediction, too. 

\begin{figure}
        \begin{center}
                \includegraphics[scale=0.6]{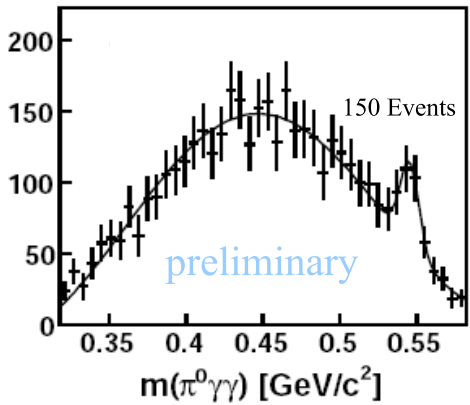}
                \caption{Invariant mass distribution for selected $\pi^0 \gamma \gamma$ events. Roughly 150 $\eta \to \pi^0 \gamma \gamma$ events sit on a big backgrounds, stemming mainly from $\eta \to 3\pi^0$ decays and $\pi^0 \pi^0$ photoproduction. The solid line shows a fit of a gaussian for the $\eta$ peak plus a polynomial for the background.}
                \label{pspi0gg}
        \end{center}
\end{figure}

Figure~\ref{pspi0gg} shows the invariant mass for selected $\pi^0 \gamma \gamma$ events measured at MAMI. The 150 $\eta$ events sitting on a big background are by far not sufficient to produce an informative Dalitz plot. Also this statistics is lower than in the recently published result of the Crystal Ball at AGS~\cite{Pra08}, but showed a much better signal to background ratio. Nevertheless, a preliminary branching ratio $BR(\eta \to \pi^0 \gamma \gamma) = (2.25 \pm 0.46_{stat} \pm 0.17_{syst}) \cdot 10^{-4}$ was determined. The Crystal Ball at MAMI results agrees with the value obtained at BNL within one standard deviation. Both are presently the most precise data for this decay.

\subsection{Future Perspectives in $\eta$ Physics}\label{hdetafut}

After improving several setup components (PbWO$_4$ crystals in TAPS, faster read--out electronics) a new $\eta$ physics program will be launched in 2010. Within 800 hours of beam time $2.5 \cdot 10^8$ $\eta$ mesons will be produced, giving the most precise results for many $\eta$ decays. Six neutral decay modes of the $\eta$ meson will be investigated simultaneously. The emphasis will be on the measurement of the Dalitz plot and decay spectrum of $\eta \to \pi^0 \gamma \gamma$, which was already preliminary investigated (see above). About 10000 useful $\eta \to \pi^0 \gamma \gamma$ events are expected. Furthermore, the Dalitz plot of $\eta \to 3\pi^0$ will be measured to investigate the speculated quadratic term for $R(z)$ and the impact of the cusp at the opening of $\pi^+ \pi^- \to \pi^0 \pi^0$ exchange reaction on the Dalitz plot parameter. More than $20 \cdot 10^6$ $\eta \to 3\pi^0$ events are expected, improving the old MAMI statistics by roughly one order of magnitude. For many $C$-- and $CP$--violating $\eta$ decays only weak upper limits for the branching ratios exist. The limits for the $C$--violating $\eta \to \pi^0 \pi^0 \gamma$, $\eta \to \pi^0 \pi^0 \pi^0 \gamma$ and $\eta \to 3\gamma$ decays, and the $CP$--violating $\eta \to 4\pi^0$ decay will be improved by a factor 10 with the expected $2.5 \cdot 10^8$ $\eta$ mesons produced.

\section{$\eta'$ Physics}\label{hdetap}

The decays of the $\eta'$ meson have been explored only with very limited statistics so far. For the most prominent neutral decay $\eta' \to \eta \pi^0 \pi^0$, with a branching ratio of $(20.7 \pm 1.2 )$~\%, the GAMS collaboration very recently published a result based on $15 \cdot 10^3$ events~\cite{Bli09}. This represents currently the highest available statistics for this decay. For the $\eta' \to 3\pi^0$ decay with a branching ratio of ($1.54 \pm 0.26) \cdot 10^{-3}$ the GAMS collaboration presented the worlds highest statistics with $235 \pm 45$ events \cite{Bli08}. It is very important to further investigate the $\eta'$ decays with high statistics. Especially, existing experimental data on $\eta'$ photoproduction are rather scarce.

Up to now, only tests concerning the $\eta'$ meson have been done with the Crystal Ball, because until the end of 2006 the maximum electron energy available at MAMI was $E_0 = 883$~MeV, while the $\eta'$ photoproduction threshold is $E_{thr} \approx 1447$~MeV. In September 2009 a further increase of the MAMI beam energy was achieved and now the maximum electron energy lies at $E_0 = 1604$~MeV. At that value the $\eta'$ photoproduction cross section reaches a plateau. First test beams at $E_0 = 1508$~MeV and $E_0 = 1558$~MeV showed that $\eta'$ mesons can be identified in a clean experimental environment at MAMI, but so far the statistics was not sufficient to obtain results with improved accuracies. To change this, a new tagging device (end--point tagger) is under construction (see sect.~\ref{hdend}), which will significantly improve the possibilities for $\eta'$ physics at MAMI. Upon completion of this apparatus a huge physical program concerning the $\eta'$ will be investigated (see sect.~\ref{hdetapfut}).

\subsection{End--point Tagger}\label{hdend}

Since the $\eta'$ photoproduction threshold is at $E_{thr} \approx 1447$\,MeV, almost the full photon energy range accessible at MAMI for $\eta'$ production is not covered by the Glasgow photon tagging spectrometer (see above). Therefore, a new tagging device (end--point tagger) is currently constructed that will detect post--radiating electrons with energies between $E_{e^-} = 10$\,MeV and $E_{e^-} = 150$\,MeV. Under the defined conditions these energies correspond to photon energies between $E_\gamma = 1548$\,MeV and $E_\gamma = 1408$\,MeV. With the new maximum energy $E_0 = 1604$\,MeV photon energies between $E_\gamma \approx 1590$\,MeV and $E_\gamma \approx 1450$\,MeV could be tagged with the end--point tagger. Thus, at least 100\,MeV above the $\eta'$ photoproduction threshold will be covered.

To study the neutral decays of the $\eta'$ meson, the end--point tagger is essential for a clean trigger. So far, $\eta'$ mesons were already measured with the Crystal Ball at MAMI, but the identification relied on the detection of outgoing protons, which mostly hit TAPS. TAPS is suitable for measuring photons, but the identification of protons is not very efficient. With the end--point tagger the detection and identification of the protons would not be required any more, and, in addition, the beam rate could be increased due to a more precise trigger.

\subsection{Future Perspectives in $\eta'$ Physics}\label{hdetapfut}

With the installation of the end--point tagger and several setup improvements (PbWO$_4$ crystals in TAPS, faster read--out electronics) one of the first high precision $\eta'$ programs in the world will be started. Approximately $15 \cdot 10^3$ produced $\eta'$ mesons per hour are expected. If the new maximum energy of $E_0 = 1604$~MeV can be run reliably this rate will be raised by roughly another factor two. Then all results on $\eta'$ decay channels investigated at MAMI will outreach precisions of older experiments by at least one order of magnitude.

The emphasis at MAMI will be on the investigation of the main neutral decay channels of the $\eta'$ meson. Among these is the decay $\eta' \to \eta \pi^0 \pi^0$. The analysis of the $\eta' \to \eta \pi^0 \pi^0$ decay amplitude is hampered by the occurrence of a cusp effect in the $\pi^0 \pi^0$ invariant mass spectrum, due to the opening of the $\pi^+ \pi^- \to \pi^0 \pi^0$ charge--exchange reaction. In~\cite{Kub09} a formalism, using $\eta' \to \eta \pi^0 \pi^0$ decays, has been developed to extract the $\pi\pi$ scattering lengths $a_0$ and $a_2$ from a detailed study of the shape of the Dalitz plot and the cusp. The authors of~\cite{Kub09} predict a cusp effect of the order of 8\,\%. Thus, this decay offers the possibility to study the $\pi \pi$ scattering length from the shape of the Dalitz plot, and to investigate $\pi \eta$ interactions. Because the energy release in the $\eta' \to \eta \pi^0 \pi^0$ decay is small, only 141~MeV, the matrix element can be described by the following formulae:
\begin{equation}
        |M|^2 = |1 + \alpha y |^2  + d x^2
\end{equation}
or
\begin{equation}
	|M|^2 = 1 + a y + b y^2 + d x^2,
\end{equation}
where $x$ and $y$ are the Dalitz variables and
\begin{eqnarray}
        a = 2 Re(\alpha) & \mathrm{and} & b = Re^2(\alpha) + Im^2(\alpha).
\end{eqnarray}
For the $\eta' \to \eta \pi^0 \pi^0$ decay Bose--Einstein symmetry forbids a linear term in $x$. A non--vanishing term $c x$ for the $\eta' \to \eta \pi^+ \pi^-$ channel would indicate $C$--violation. The currently most precise analysis of the neutral channel used $15 \cdot 10^3$ events. At MAMI approximately $4 \cdot 10^5$ $\eta' \to \eta \pi^0 \pi^0$ useful events will be measured, improving results by more than one order magnitude.

Another neutral decay that will be examined is $\eta' \to 3\pi^0$. The $\eta' \to 3\pi^0$ decay has a larger phase space than the $\eta' \to \eta \pi^0 \pi^0$ channel, but is G--forbidden. It still occurs as a consequence of the G--violating mass term of the QCD Lagrangian with BR($\eta' \to 3\pi^0)=(1.54 \pm 0.26) \cdot 10^{-3}$. The physics of the $\eta' \to 3\pi^0$ decay is similar to $\eta' \to \eta \pi^0 \pi^0$ and $\eta \to 3\pi^0$. Due to final state rescattering of the pions an energy dependence on the shape of the Dalitz plot is introduced. The matrix element for this decay can be parametrised similar to eq.~\ref{eqAmpExpand}. So far, this decay has only been observed by the GAMS collaboration based on $235 \pm 45$ events. At MAMI several thousand events of this decay are expected.

Furthermore, it is planned to measure the branching ratio for $\eta' \to \gamma \gamma$. This ratio, in conjunction with $\mathrm{BR}(\eta \to \gamma \gamma)$, allows to investigate the $SU(3)$ singlet--octet mixing angle. Because each of the physical states, $\eta$ and $\eta'$, has both singlet and octet components, in a more evolved parametrisation two mixing angles can be introduced and one defines four independent decay constants. Using this approach, it is possible to calculate the decay widths for the $\eta \to \gamma \gamma$ and $\eta' \to \gamma \gamma$ decays, as shown in~\cite{Esc05}. In this calculation four unknown parameters ($\theta_8, \theta_0, f_8, f_0$) appear. To determine these values one needs additional experimental constraints as the decay widths of the $\eta \to \gamma \gamma$ and $\eta' \to \gamma \gamma$ decays. By measuring these two decays and improving the accuracies of the widths we can give an important input for the determination of the mixing angles $\theta_8$ and $\theta_0$. The uncertainty of the width of the $\eta' \to \gamma \gamma$ decay, which is of the order of 10\,\%, is currently determined by both the uncertainty of the branching ratio and the uncertainty of the total width of the $\eta'$. With the expected 40000 events in 600 hours of data taking, we will decrease the uncertainty of BR$(\eta' \to \gamma \gamma)$ greatly.

Recently the KLOE collaboration published data on the $\eta$--$\eta'$ mixing angle and a possible gluonium content to the $\eta'$~\cite{Amb07b,DiM08} determined from $R = \Gamma(\phi \to \eta \gamma)/\Gamma(\phi \to \eta' \gamma)$. They found a non--zero gluonium content to the $\eta'$, contrary to the theoretical calculation in~\cite{Esc07}. Further investigations of the $\eta$--$\eta'$ mixing scheme are needed and precise data from experiments with the Crystal Ball at MAMI will give important input to these investigations.

Also in the field of $\eta'$ mesons  symmetry violating decays will be investigated at MAMI. The $\eta' \to \eta e^+ e^-$, $\eta' \to \pi^0 e^+ e^-$ decays are forbidden by $C$--invariance in first order, but allowed in second. Another test of $C$--invariance is the branching ratio of the decay $\eta' \to 3\gamma$. The $\eta' \to 4\pi^0$ decays violates $P$-- and $CP$--invariance. Lepton number violation can be studied in $\eta' \to e \mu$. For all these decays it will be possible, with the expected count rates, to improve the upper limits of the branching ratios by more than one order of magnitude.

\section{Conclusions}

This proceeding described recent results from the Crystal Ball at MAMI experiment in $\eta$ physics. A new value for the $\eta$ meson mass was determined, the Dalitz plot parameter was extracted from the worlds greatest $\eta \to 3\pi^0$ event samples, and a new branching ratio of the $\eta \to \pi^0 \gamma \gamma$ decay was found. In near future very interesting and important results on $\pi \pi$ scattering lengths, cusp effects, Dalitz plot parameters and symmetry violations are to be expected from high precision measurements on $\eta$ and $\eta'$ mesons at one of the leading facilities in this field, the Crystal Ball at MAMI.\\

{\bf Acknowledgments:} The author wishes to thank the accelerator group of MAMI for the precise and very stable beam conditions. This work was supported by the German state Rhineland-Palatinate (EMG ``Elementarkr\"afte und mathematische Grundlagen''), the Deutsche Forschungsgemeinschaft (SFB 443, SFB/TR 16), the DFG-\-RFBR (Grant No. 05-02-04014), European Community-Research Infrastructure Activity under the FP6 ``Structuring the European Research Area programme" (HadronPhysics, Contract No. RII3-CT-2004-506078).

\end{document}